
\baselineskip 22pt
\font\prlsize=cmr12
\prlsize

\centerline {\bf Semiquantal dynamics of fluctuations:}
\centerline {\bf ostensible quantum chaos }

\bigskip \centerline{Arjendu K. Pattanayak and William
C. Schieve} \medskip
\centerline{Prigogine Center for Statistical Mechanics and Complex Systems,}
\centerline{The University of Texas, Austin, Texas 78712}
\bigskip
\bigskip

\smallskip
{\bf Abstract}

 {\narrower \narrower \medskip
The time-dependent variational principle using generalized Gaussian trial
functions yields a finite dimensional approximation to the full
quantum dynamics and is used in many disciplines.
It is shown how these 'semi-quantum' dynamics may be derived via
the Ehrenfest theorem and recast as an extended classical gradient system
with the fluctuation variables coupled to the average variables.
An extended potential is constructed for a one-dimensional system.
The semiquantal behavior is shown to be chaotic even though the system
has regular classical behavior and the quantum behavior had been assumed
regular. \smallskip}
\smallskip
PACS number(s): 05.45.,03.65.S,05.40.
\bigskip

There has been substantial effort made, over the years, to understand
quantum systems using a system of few classical
variables. These can be motivated in the $\hbar \to 0$ limit,
such as effective potential techniques[1,2] and  semi-classical methods
like that of WKB, which lead to the Einstein-Brillouin-Keller (EBK)
[3] quantization rules for integrable systems. These also include
approximations to the Feynman path-integral formulation [4],
used to derive the Periodic-Orbit Trace Formula for chaotic systems
[5]. This relates the spectrum of the quantum system to a weighted
sum over the unstable periodic orbits of the classical system.
They can also arise, as in Quantum Chromo Dynamics for example [6], in the
limit of large $N$ (number of degrees of freedom), and a classical phase
space can be shown to exist in the $N=\infty$ limit [7].
 Further, there are many equivalent mean-field theories that are used
in nuclear physics, quantum chemistry, quantum field theory[8,9,10], condensed
matter, statistial mechanics and optics (see the excellent reviews [11,12]).
These are known variously as the time-dependent Hartree-Fock method, the
Gaussian variational approximation, etc.
One way to understand these approaches is to consider the time-dependent
variational principle (TDVP) formulation [8,13,14]
wherein one posits the action
$$
\Gamma = \int dt < \Psi,t| i\hbar {\partial \over \partial t} -
H| \Psi, t>.
\eqno(1)
$$
The requirement that $\delta \Gamma = 0$ against independent variations of
$<\Psi,t|$, and $|\Psi,t>$ yields the Schr\"odinger equation and its
complex conjugate as the respective Euler-Lagrange equations
(note that a ray rotation, i.e. $| \Psi, t> \to exp(i\lambda (t)/\hbar)
| \Psi, t>$ leaves the variational equations unchanged).
The true solution may be approximated by restricting
the choice of states to a subspace of the full Hilbert space
and finding the path along which $\delta \Gamma = 0$ within this subspace.
In exceptional circumstances, this restriction
is the true space of the problem and the solution is then exact.
If we restrict $| \Psi, t>$ to the family of coherent
states[11,12,15-17] this corresponds, in ordinary quantum mechanics
to a wave-packet of the form
$$
\Psi (x) = N(\pi \hbar )^{n/4}exp[1/\hbar (-{1\over 2}({\bf x - q})^2
+ i{\bf p.(x-q)} + i \lambda)].
\eqno(2)
$$
Here $N ,\lambda$ are the normalization and phase respectively and ${\bf q,p}$
specify a point in the classical phase space of dimension $2n$. It is
straightforward to show [18] that to ${\cal O}(\hbar)$, the equations for this
system are Hamilton's equations for ${\bf p, q}$,
conserve the norm $N$ and yield
$$
\dot \lambda = {\bf p.\dot q} - H({\bf q,p}),
\eqno (3)
$$
where $H$ is the classical Hamiltonian. Hence, $\lambda(t)$ is equal
to the classical action; this can be used
[18] to identify the semiclassical Bohr-Sommerfeld phase with the
Aharanov-Anandan form of Berry's phase[19].
A more interesting situation arises when, as below, we consider
squeezed coherent states (the wave-packets are allowed to spread).
Along with the dynamics for the centroid of the wave-packet,
we now also consider equations of motion for its spread.
The full equations give us the mean-field theories mentioned above[9-12,20].
If, alternatively, in the Taylor series expansions around
the centroid, we truncate to second order in derivatives
of the Hamiltonian, we recover the equations of Heller[21], who has used
this method extensively in studying 'quantum chaos'.

Implicit in these 'semi-quantum' methods is the assumption that these dynamics
are qualitatively similar to those of
either the fully classical limit or the full quantum system. If the classical
behavior is integrable, the semiquantal dynamics are said not to break the
integrability[22]. Consideration of quantal fluctuations is supposed
to 'supress' chaos and the full quantum dynamics are usually said to be
regular.
In this Letter, we demonstrate that this is {\sl not}
always true. Using the example of the double-well potential in ordinary
quantum mechanics, we show that quantal effects, in fact, {\sl induce} chaos.
While neither the quantum nor the classical system are understood to
display the sensitive dependence to initial conditions characteristic of
chaos, the semiquantal dynamics does do so.
We justify here the use of the terminology 'semiquantal'. This
is not a trivial point - we derive the dynamics directly from the quantum
Hamiltonian with no reference to the classical limit.
In fact, it can be shown[22] that semiquantal dynamics exist
even for systems without a well-defined classical dynamics.
However, if the traditional 'semiclassical' and classical limits do exist,
they can be recovered as special limits of semiquantal dynamics.
The terminology is hence most apposite.

While the standard approach uses the TDVP,
we provide here an alternative derivation of
semiquantal dynamics via the Ehrenfest theorem. We believe that this
is somewhat more intuitive and is similar to formulations
in non-equilibrium statistical mechanics. The dynamics obtained
are the same as in the TDVP approach. Consider
a particle of unit mass moving in a one-dimensional time-independent
bounded potential with a Hamiltonian $ \hat H = {\hat p ^2 \over 2} +
V(\hat x)$ where ${\hat O}$ denotes operators. The equations of
motion for the centroid of a wavepacket
representing the particle are
$$
{d \over dt} < \hat x> = <\hat p>,
\eqno(4a)
$$
$$
{d \over dt} <\hat p> = - <{\partial V (\hat x ) \over \partial \hat x}>
\eqno(4b)
$$
 where the $<>$ indicate expectation values. In general
the centroid does not follow the classical trajectory.
We now expand the equations around the centroid using the identity
$$
<F(\hat u)> = {1 \over n!} <\hat U^n > F^{(n)},\ \ \   n \geq 0
\eqno(5)
$$
where
$F^{(n)} = {\partial ^n F / \partial u^n} |_{<\hat u>}$
and $\hat U = \hat u - <\hat u>$ (the repeated index summation convention
is used throughout, unless otherwise specified).
Using this and operator commutation rules, we can generate a countably
infinite number of moment equations (corresponding to the infinite dimensional
Hilbert space of the problem).
The assumption that the wavepacket is a squeezed coherent
state renders the space finite; it provides the relations
$<\hat X^{2m}> = { (2m)! \mu^m \over m! 2^m} {\rm( no\ summation)} ,
<\hat X ^{2m+1}> =0 , 4\mu <\hat P^2> =   \hbar^2 + \alpha^2$ and$
<\hat X \hat P + \hat P \hat X> =  \alpha,
$
which are easily recognised as those for generalized Gaussian
wave-functions[9-12,20].(This assumption is precisely that of
the TDVP: the wavepacket is restricted to a given subspace).
This yields the equations
$$ \eqalignno {
{d x \over dt}  &= p
&(6a)\cr
{d p \over dt} &= - {\mu ^m \over m! 2^{m}} V^{(2m+1)} (x), \ \ \ m= 0, 1,
 \ldots
& (6b)\cr
{d \mu \over dt} &=  \alpha , &(6c)\cr
{d \alpha \over dt} &= { \hbar^2 + \alpha^2 \over 2 \mu}
-{\mu^m \over (m-1)!2^{m-2}}V^{(2m)}(x). \ \ \ m =1,2 \ldots
&(6d)\cr
}$$
The system is now reduced to the dynamics of $x, p,
\mu$ and $\alpha$ ( where we write $x,p$ for $<\hat x>,<\hat p>$)
and are exactly those derived from the
action principle [23].

We now introduce the change of variables $\mu = \rho^2,$ and
$\alpha = 2 \rho \pi,$
transforming equations (6) to
$$ \eqalignno {
{d x \over dt} &= p ,
&(7a)\cr
{d p\over dt}  &=
 - {\rho ^{2m} \over m! 2^m} V^{(2m+1)} (x), \ \ \ m= 0, 1, \ldots
& (7b)\cr
{d \rho \over dt}  &=  \pi, &(7c)\cr
{d \pi\over dt} &= { \hbar^2  \over 4 \rho ^3}
-{\rho^{2m-1} \over (m-1)!2^{m-1}}V^{(2m)}(x), \ \ \ m =1,2 \ldots
&(7d)\cr
}$$
Remarkably, these new variables form an explicit
canonically conjugate set, yielding a {\sl classical Hamiltonian
phase space} as our approximation to the Hilbert space.
The classical degrees of freedom are the 'average' variables
$x,p$ and the 'fluctuation' variables $\rho, \pi$, respectively; the
associated Hamiltonian is
$$
H_{ext} = {p^2 \over 2} + { \pi ^2 \over 2 } + V_{ext}(x,\rho);
\eqno(8a)
$$
$$
V_{ext} (x,\rho) = V(x) + {\hbar ^2\over 8 \rho^2} +
{\rho^{2m} \over m!2^m }V^{(2m)}(x), \ \ \ m =1,2 \ldots
\eqno(8b)
$$
 where the subscript {\it ext} indicates the 'extended'
potential and Hamiltonian.
This formulation is very interesting
and possibly quite powerful; it provides us with an explicit gradient system,
and the extended potential provides a simple visualization of
the geometry of the semiquantal space.
We may thus get a qualitative feel for the semiquantal
dynamics before proceeding to detailed (numerical) analysis.
We note here that
a) both the fluctuation and average variables are treated on the same footing
and the phase space is dimensionally consistent: $\rho$ has the dimensions
of length and $\pi$ that of momentum,
b) the value of $H_{ext}$ is $<\hat H>$ under this approximation,
and is conserved, c) $V_{ext}$ has an infinite barrier at $\rho = 0$, hence
'quantum fluctuations' can not be zero except in the
limit $\hbar \to 0$ and
d) $V_{ext}$ is symmetric in $\rho$
corresponding to a choice of sign in $\rho =\sqrt \mu = \sqrt {<X^2>} $;
the infinite barrier renders the choice one
of convenience and of no physical significance.

We note also that this formulation is exact for the simple harmonic oscillator.
For this system, the variables $(x,p)$ decouple from $(\rho, \pi)$.
The average variables hence execute the usual classical harmonic motion. The
fluctation variables are a bound one-degree-of-freedom problem, and in
general execute oscillatory motion. It is possible to find a fixed point
in these variables, however. This is the familiar example of a Gaussian
wavepacket that executes harmonic motion with a fixed spread (see Ref. 21).
The extended potential, in this case, provides us with the quantum corrections
due the 'zero-point fluctuations'.

We consider now the dynamics of
the simple double-well system with the Hamiltonian
$\hat H = {\hat p ^2 \over 2} - {a \hat x ^2 \over 2}
+ { b \over 4} \hat x ^4.
$
The extended Hamiltonian is
$$
H_{ext} = {p^2 \over 2} + { \pi ^2 \over 2 } -{a \over 2} (x^2 + \rho ^2)
+ {1 \over 8 \rho ^2} + {b \over 4}(x^4 + 3\rho ^4 + 6x^2 \rho ^2 )
\eqno(9)
$$
where we have set $\hbar =1$.
If we examine
$V_{ext}$ in the upper half of the $x-\rho$ plane,
it is straightforward to show that there is always a minimum
for the potential at $(x,\rho)=(0,\bar \rho)$, where $\bar \rho$ is the
largest positive real solution of
$
a\rho^4 - 3b\rho ^6 + 1/4 = 0.
$
For $\xi = {32a^3 \over 243 b^2} \leq 1 $, this is the
only minimum; quantum effects are so large that the the well barrier is
effectively absent.
As $\xi$ increases, two minima appear in the half-plane, corresponding
to the well minima of the original problem; two saddle-points also emerge,
interestingly.
Also, the $(x,p)$ system as driven by the $(\rho,\pi)$ system
is like a non-linearly driven Duffing oscillator[24] with back reaction.
With these ingredients, it is not surprising that the equations
$$ \eqalignno {
{d x \over dt} &= p ,
&(10a)\cr
{d p \over dt} &= ax -bx^3 -3bx\rho ^2,
& (10b)\cr
{d \rho \over dt} &=  \pi ,
&(10c)\cr
{d \pi \over dt} &= {1 \over 4\rho ^3}  + \rho(a - 3bx^2)  -3b\rho ^3
&(10d)\cr
}$$
(which are the just the explicit form of equation 7
with $\hat H(\hat p ,\hat x)$ as above)
exhibit chaos. We have numerically investigated these equations over a range
of parameter values and initial conditions.
We summarize here some of these results at the typical parameter values
of $a=1,b=.01$ i.e., where quantum effects are small but not negligible.
At low energies, as is typical for Hamiltonian systems,
there are only integrable orbits. As the energy increases,
chaotic orbits also emerge; the two types are very near
each other in choice of initial conditions. The details of the phase-space
structure will not be discussed here. However, the existence
of chaos in this system has been established
using Poincar\'e sections and Lyapunov exponents calculations[25].
In Figs.1,2 we show a typical Poincar\'e section; taken with
$\dot \rho= 0, \ddot \rho \ge 0$.
We can see the features characteristic of
Hamiltonian chaos: the typical stochastic web
structure, and in Fig.2, which is an enlargement of Fig.1,
the familiar appearance of islands in this stochastic sea [26].
The associated largest Lyapunov exponent $\lambda_{max}$ equals $0.125$.
The calculation of Lyapunov exponents is also typical, with a slow saturation,
and a residual oscillation
of about $5\%$. The exponents are symmetric around $0$ and hence sum
to $0$ (to within $10^{-6}$).
We note that for various orbits (as for this one) where the classical motion
would have remained confined to one well, the fluctuation dimension
provides a means for the system to 'tunnel' between the two wells.

In the classical limit, this system has only two independent
first-order differential equations, insufficient for chaos.
The quantum dynamics are said to have 'no chaos', based on the fact
that Schrodinger's equation is a linear partial differential equation;
the dynamics are said to be formally equivalent to a classical harmonic
oscillator system in quadrature on a $2N$-dimensional classical phase space.
It has been recently argued[27] that this is too simplistic: the
constraint of square integrability and arbitrary ray rotation renders the space
a compact complex projected space $CP(N-1)$ which has a different topological
structure and may display substantially different behavior.
There is a conjecture[22], further, that
in systems with inherently infinite-dimensional Hilbert spaces, semiquantal
effects enhance chaos (see Ref. 22 for details).
For this system, the dynamical effect of the fluctuation variables
is to {\sl induce } chaos. This may be regarded as support for the
arguments of Ref. 22; if, in fact, the full quantum system has 'no chaos',
this is anomalous behavior that exists only in the semiquantal limit.

In summary, we have introduced above a way of viewing the
time-dependent Hartree-Fock method and other mean-field theories - the extended
potential approach. Using this formulation for the double-well system,
we have demonstrated that quantum fluctuations may induce
chaotic semiquantal behavior, in keeping with a recent conjecture.
Irrespective, care is indicated[28]: while variational and other classical
approximations made to quantum systems are quite powerful,
their qualitative behavior may be anomalous
and may not exist in the full quantum system nor its completely classical
limit.

AKP acknowledges the Robert A. Welch Foundation (Grant
No. F-0365) for partial support.

\leftline {\bf References}
\medskip
\item {1.} P. Stevenson, a){\sl Phys. Rev. D} {\bf 30}, 1712 (1984);
 b){\sl ibid} {\bf 32}, 1389 (1985) and references therein.
\smallskip
\item {2.}a) A.K. Pattanayak and W.C. Schieve,
{\sl Phys. Rev. A} {\bf 46}, 1821 (1992); b) L. Carlson and W.C. Schieve,
{\sl Phys. Rev. A} {\bf 40}, 5896 (1989).
\smallskip
\item {3.} I.C. Percival, {\sl Adv. Chem. Phys. }{\bf 36}, 1 (1977)
provides a review of this work.
\smallskip
\item {4.}L.S. Schulman, 'Techniques and Applications of Path Integration'
(Wiley-Interscience, 1981).
\item {5.}M. Gutzwiller, 'Chaos in Classical and Quantum Mechanics',
(Springer-Verlag ,1990).
\smallskip
\item {6.}E. Witten,{\sl Nucl. Phys. }{\bf B160}, 57 (1979).
\smallskip
\item {7.}L.G. Yaffe, {\sl Rev. Mod. Phys. }{\bf 54}, 407 (1982).
\smallskip
\item {8.}R. Jackiw and A. Kerman, {\sl Phys. Lett. }{\bf 71A}, 158 (1979).
\smallskip
\item {9.}F. Cooper, S.-Y. Pi and P.N. Stancioff,
{\sl Phys. Rev. D} {\bf 34}, 3831 (1986).
\smallskip
\item {10.}A. Kovner and B. Rosenstein,
{\sl Phys. Rev.} {\bf D39}, 2332, (1989).
\smallskip
\item {11.}J. Klauder and B.-S. Skagerstam, 'Coherent States: Applications
in Physics and Mathematical Physics' (World Scientific, 1985).
\smallskip
\item {12.}W.-M. Zhang, D.H. Feng and R. Gilmore,
{\sl Rev. Mod. Phys. }{\bf 62}, 867 (1990).
\smallskip
\item {13.}P.A.M. Dirac, Appendix to the Russian edition of 'The
Principles of Quantum Mechanics', as cited by Frenkel, I.I., 'Wave Mechanics,
Advanced General Theory' (Clarendon Press, Oxford, 1934) (pg. 253, 436).
\smallskip
\item {14.}J. Klauder,{\sl J. Math. Phys.},{\bf 4}, 1055 (1963); 1058 (1963).
\smallskip
\item {15.}E.C.G. Sudarshan, {\sl Phys. Rev Lett.}{\bf 10},277 (1963)
\smallskip
\item {16.}R. Glauber, {\sl Phys. Rev. Lett. }{\bf 10},84 (1963);
{\sl Phys. Rev. }{\bf 130}, 2529 (1963).
\smallskip
\item {17.}A.M. Perelmov, 'Generalized Coherent States and their Applications'
 (Springer-Verlag, 1986)
\smallskip
\item {18.}R.G. Littlejohn, {\sl Phys. Rev. Lett. }{\bf 61}, 2159 (1988).
\smallskip
\item {19.}M.V. Berry,{\sl Proc. Roy. Soc.} {\bf A392}, 45 (1984);
A. Aharanov and J. Anandan, {\sl Phys. Rev. Lett.} {\bf 56}, 2000 (1986).
\smallskip
\item {20.}Y. Tsue, {\sl Prog. Theor. Phys. }{\bf 88}, 911 (1992)
and references therein.
\smallskip
\item {21.}E.J. Heller, {\sl J. Chem. Phys.} {\bf 62}, 1544 (1975);
in 'Chaos and Quantum Physics', Proceedings of the
Les Houches Summer School 1989, (North-Holland, 1991).
\smallskip
\item {22.} W.-M. Zhang and D.H. Feng , 'Quantum Nonintegrability',
 {\sl Phys. Rep.} (In press, 1994); {\sl Mod. Phys. Lett}{\bf A8},1417 (1993).
\smallskip
\item {23.}To see the equivalence, equate our variables
$x, p, \mu,$ and $\alpha$  with $q, p, \hbar G,$  and  $4 \hbar G \Pi$
respectively of Refs 8,9 and 20.
\smallskip
\item {24.}J. Guckenheimer and P. Holmes, 'Non-linear Oscillations,
Dynamical Systems and Bifurcations of Vector Fields' (Springer-Verlag, 1983).
\smallskip
\item {25.}A. Wolf, J.B. Swift, H.L. Swinney and J.A. Vastano,
{\sl Physica } {\bf 16D}, 285 (1985).
\smallskip
\item {26.}G.M. Zaslavsky, R.Z. Sagdeev,D.A. Usikov and A.A. Chernikov,
'Weak Chaos and Quasi-regular Patterns' (Cambridge University Press, 1991).
\smallskip
\item {27.}A. Heslot, {\sl Phys. Rev. D } {\bf 31}, 1341 (1985).
\smallskip
\item {28.}See R. Balian and M. V\'en\'eroni,{\sl Ann. Phys.}
(N.Y.) {\bf 87}, 29 (1988) for similar comments in a different context.
\vfill
{\bf Figure Captions}
\item {1.} Fig.1. Poincar\'e sections in the $x,p$ plane at an energy of
$H_{ext} = -1.25$. There are about 30000 points shown. $\lambda_{max}= 0.125$.
\item {2.} Fig.2. Enlargement of Fig.1. Notice the distinctive stability
islands in the stochastic sea.
\vfill
\eject

\end